\documentclass[
	,reprint
	,preprintnumbers
	,floatfix
	,nofootinbib
	,amsmath
	,amssymb
	,aps
	,pra
	,letterpaper
]{revtex4-1}
\usepackage{graphicx}
\usepackage{mathtools}
\usepackage{xcolor}
\usepackage[
]{physics}

\usepackage{mleftright}\mleftright
\usepackage[T1]{fontenc}
\usepackage[utf8]{inputenc}
\usepackage{lmodern}
\usepackage{isomath}
\usepackage{upgreek}
\usepackage{libertineRoman}
\usepackage[
	,slantedGreek
	,libertine
	,libaltvw
	,liby
]{newtxmath}
\usepackage{bm}
\usepackage[
	,unicode
	,colorlinks
	,allcolors=blue!90!cyan!80!black!90!white
	,setpagesize=false
	,bookmarksnumbered=true
	,bookmarksopen=true
	,pdfborder={0 0 0}
]{hyperref}
\usepackage{bookmark}
\allowdisplaybreaks[0]
\newcommand{\br}{\bm{r}}
\newcommand{\bp}{\bm{p}}
\newcommand{\bb}{\bm{b}}
\newcommand{\bv}{\bm{v}}
\newcommand{\bu}{\bm{u}}
\newcommand{\bhz}{\mathbf{\hat z}}
\newcommand{\cI}{\mathcal{I}}

\newcommand{\veps}{\varepsilon}
\newcommand{\zomega}{\omega}
\newcommand{\ui}{\mathrm{i}}
\newcommand{\ue}{\mathrm{e}}
\newcommand{\xl}{\lambda}
\newcommand{\xa}{\alpha}

\newcommand{\trap}{{\mathrm{trap}}}

\newcommand{\mix}{\mathrm{in}}
\newcommand{\F}{\mathrm{F}}

\newcommand{\B}{\mathrm{B}}
\newcommand{\BB}{\mathrm{BB}}
\newcommand{\BF}{\mathrm{BF}}
\newcommand{\FB}{\mathrm{FB}}
\newcommand{\BEC}{\mathrm{BEC}}

\newcommand{\TF}{T_\mathrm{F}}

\newcommand{\fast}{\mathrm{fast}}
\newcommand{\slow}{\mathrm{slow}}

\newcommand{\kB}{k_{\mathrm B}}

\begin{document}
\title[Dipole oscillation]{%
Dipole oscillation of a trapped Bose--Fermi-mixture gas
\\ 
in collisionless and hydrodynamic regimes
}
\author{%
Yoji~Asano,
Shohei~Watabe, and 
Tetsuro~Nikuni
}
\affiliation{%
Department of Physics,
Tokyo University of Science,
1-3 Kagurazaka,
Shinjuku-ku,
Tokyo,
162-8601,
Japan
}
\date{\today}
\begin{abstract}
Dipole oscillation is studied in a normal phase of a trapped Bose--Fermi-mixture gas composed of single-species bosons and single-species fermions.
Applying the moment method to the linearized Boltzmann equation, we derive a closed set of equations of motion for the center-of-mass position and momentum of both components.
By solving the coupled equations, we reveal the behavior of dipole modes in the transition between the collisionless regime and the hydrodynamic regime.
We find that two oscillating modes in the collisionless regime have distinct fates in the hydrodynamic regime; one collisionless mode shows a crossover to a hydrodynamic in-phase mode, and the other collisionless mode shows a transition to two purely damped modes.
The temperature dependence of these dipole modes are also discussed.
\end{abstract}
\maketitle

\section{Introduction}
\label{sec:introduction}
The technology of ultracold atomic gases has allowed excellent opportunities for the experimental study of quantum many-body systems.
One of the most interesting achievements is the realization of quantum mixtures of gases of two
isotopes~\cite{schreck_2001_PRA_64_sympathetic,fukuhara_2007_PRA_76_bose,lu_2012_PRL_108_quantum}
and two different species of
atoms~\cite{ferlaino_2003_JOBQSO_5_dipolar,fukuhara_2009_APB_96_quadrupole,roy_2016_PRA_94_photoassociative,desalvo_2017_PRL_119_observation,ravensbergen_2019_arXiv_strongly}.
Compared with the well-known mixture of quantum liquids, $^3\mathrm{He}$ and $^4\mathrm{He}$, ultracold atomic gases have a richer variety of combinations, including changes in quantum statistics, number of particles, mass of the particles, interaction strength, and geometry of
the confinement~\cite{onofrio_2016_PU_59_physics}.

Recent interest has focused on Bose--Fermi mixtures.
In particular, the static properties of heteronuclear Bose--Fermi
molecules~\cite{roy_2016_PRA_94_photoassociative}
and degenerate Fermi gases with a Bose--Einstein condensate
(BEC)~\cite{desalvo_2017_PRL_119_observation}
have been investigated.
Some of the dynamics, such as dipole
oscillations~\cite{ferlaino_2003_JOBQSO_5_dipolar,delehaye_2015_PRL_115_critical},
quadrupole
oscillations~\cite{fukuhara_2009_APB_96_quadrupole}, 
and breathing modes of BECs with a fermionic
reservoir~\cite{huang_2019_PRA_99_breathing}
have been addressed.
Furthermore, a superfluid Bose--Fermi mixture has been created, and its center-of-mass
motion~\cite{ferrier-barbut_2014_S_345_mixture}
and critical
velocity~\cite{delehaye_2015_PRL_115_critical,castin_2015_CRP_16_la}
have been studied.
The collective modes of Bose--Fermi mixtures have been theoretically investigated, 
namely, the monopole and multipole
modes~\cite{banerjee_2007_PRA_76_collective,liu_2003_PRA_67_collisionless,vanschaeybroeck_2009_PRA_79_trapped}, 
low-lying modes in spinor Bose--Fermi mixture
gases~\cite{pixley_2015_PRL_114_damping}, 
density and single-particle
excitations~\cite{capuzzi_2001_PRA_64_zero},
and collective modes using the variational-sum-rule
approach~\cite{banerjee_2009_JPBAMOP_42_dipole,miyakawa_2000_PRA_62_sum}.

Collective modes, which exhibit fluctuation of number density, are often characterized by collisional processes, where local quantities relax to equilibrium during collisions.
In particular, the competition between two time scales is important: a relaxation time $\tau$ and the period of a collective mode, given by a frequency $\omega$.
Hydrodynamic modes are characterized by $\omega\tau\ll1$, where the relaxation time is sufficiently shorter than the oscillatory period of the collective mode.
In contrast, collisionless modes are characterized by the opposite condition $\omega\tau\gg1$.
Indeed, a smooth crossover has been observed between those collisionless and hydrodynamic modes in ultracold atomic 
gases~\cite{gensemer_2001_PRL_87_transition,buggle_2005_PRA_72_shape}.

In an earlier
study~\cite{asano_2019_JLTP_196_collective},
we examined a collective sound mode in a normal phase of a uniform Bose--Fermi mixture, composed of single-species bosons and single-species fermions interacting via $s$-wave scattering, where collisions between fermions are absent because of Pauli blocking.
Intraspecies (Bose--Bose) scattering becomes important at low temperatures close to the BEC transition temperature because of Bose enhancement, which may lead to the hydrodynamic regime.
On the other hand, Pauli blocking of fermions suppresses the interspecies (Bose--Fermi) scattering at extremely low
temperature~\cite{schreck_2003_AdP_28_mixtures}.
This suppression favors the collisionless regime.
These quantum statistical properties may give rise to interesting features of collective modes.
The earlier
study~\cite{asano_2019_JLTP_196_collective}
showed the surprising result that there exists a long-lived sound mode between the hydrodynamic regime and the collisionless regime.
This contrasts with our general knowledge, where the lifetime of a collective mode is very short in the crossover regime.

Dipole oscillation, a one-dimensional motion of the center of mass, is more accessible experimentally in ultracold gases.
An earlier experiment observed the dipole oscillation of a normal Bose--Fermi mixture in both the collisionless and hydrodynamic
regimes~\cite{ferlaino_2003_JOBQSO_5_dipolar}.
However, it is not clear how the collisionless dipole modes turn into the hydrodynamic dipole modes in mixture gases.
The moment method for solving the Boltzmann equation is useful for addressing this issue, treating both the collisionless and hydrodynamic modes in a coherent
manner~\cite{guery-odelin_1999_PRA_60_collective,nikuni_2002_PRA_65_finite,ghosh_2000_PRA_63_collective,watabe_2010_JLTP_158_zero,watabe_2010_PRA_82_dynamic,narushima_2018_JPBAMOP_51_density}.

In this paper, using the moment method we study dipole oscillation of a trapped Bose--Fermi mixture, composed of single-species bosons and single-species fermions.
We find that the dipole modes are characterized by a single relaxation time, which originates from Bose--Fermi scattering associated with conservation of the total momentum.
In the collisionless regime, two types of oscillating modes emerge.
As the relaxation time becomes short and the system enters the hydrodynamic regime, one oscillating collisionless mode shows a crossover to an oscillating in-phase mode, and the other oscillating collisionless mode disappears and turns into two purely damped modes: the fast and slow relaxation modes.
We expect that the fast relaxation mode exists in the hydrodynamic regime, since it is analogous to the well-known diffusion mode or spin drag in a uniform gas.
In a trapped gas, we find that the relative center-of-mass motion gives rise to a slow relaxing mode, where two separated clouds gradually mix with each other.

\section{Moment method}
\label{sec:method}
\subsection{Boltzmann equation}
The Boltzmann equation for the distribution function $f_\xa=f_\xa(\br,\bp,t)$, as a function of position $\br$, momentum $\bp$, and time $t$, is described by
\begin{align}&
\pdv{f_\xa}{t}+\pdv{\veps_\xa}{\bp}\cdot\pdv{f_\xa}{\br}
-\pdv{U_\xa}{\br}\cdot\pdv{f_\xa}{\bp}=\cI_\xa
\;,
\label{eq:Boltzmann_equation}
\end{align}
where the single-particle energy $\veps_\xa=\veps_\xa(\br,\bp,t)$ with atomic mass $m_\xa$ is given by 
\begin{align}&
\veps_\xa(\br,\bp,t)=\frac{\bp^2}{2m_\xa}+U_\xa
\;.
\label{eq:single-particle_energy}
\end{align}
Here, $\xa=\{\B,\F\}$ represents bosons and fermions, respectively.
The potential energy $U_\xa=U_\xa(\br,t)$ is given by a combination of the mean-field term and a trapping potential: 
\begin{subequations}
\begin{align}& 
U_\B(\br,t)=2g_\BB n_\B+g_\BF n_\F+U^\trap_\B
\label{eq:potential_energy_for_a_boson}
\;,\\&
U_\F(\br,t)=g_\BF n_\B+U^\trap_\F
\;.
\end{align}
\label{eq:potential_energy}%
\end{subequations}
Here, $g_\BB$ and $g_\BF$ are the coupling strengths of Bose--Bose and Bose--Fermi scattering, respectively, and $n_\xa=n_\xa(\br,t)$ is the number density, defined by the momentum integration of the distribution function:
\begin{align}&
n_\xa(\br,t)
=\int\!\!\!\frac{\dd[3]{p}}{(2\uppi\hbar)^3}f_\xa(\br,\bp,t)
\;.
\label{eq:number_density}
\end{align}
We note that the $s$-wave scattering lengths $a_\BB$ and $a_\BF$ are related to the coupling strengths as $g_\BB=4\uppi\hbar^2a_\BB/m_\B$ and $g_\BF=2\uppi\hbar^2 a_\BF/m_\BF$, respectively, with a reduced mass $m_\BF=m_\B m_\F/(m_\B+m_\F)$.
The Fermi--Fermi interaction is absent, because we are considering single-species fermions with $s$-wave scattering.
For the trapping potential $U^\trap_\xa=U^\trap_\xa(\br)$, we consider the axisymmetric harmonic potentials
\begin{align}&
U^\trap_\xa(\br)=\frac{m_\xa\zomega_\xa^2}{2}\bqty{\xl_\xa^2(x^2+y^2)+z^2}
\;,
\label{eq:trapping_potential}
\end{align}
where $\zomega_\xa$ and $\xl_\xa\zomega_\xa$ denote the axial and radial trap frequencies, respectively.

The collision integrals $\cI_\xa=\cI_\xa(\br,\bp,t)$ in the present case are given by
\begin{subequations}
\begin{align}&
\cI_\B
=\cI_\BB [f_\B,f_\B]+\cI_\BF[f_\B,f_\F]
\;,\\& 
\cI_\F
=\cI_\FB[f_\F,f_\B]
\;,
\end{align}
\end{subequations}
where the expression for $\cI_{\xa\beta}=\cI_{\xa\beta}[f_\xa,f_\beta]$ ($\{\xa,\beta\}=\{\B,\B\}$, $\{\B,\F\}$, $\{\F,\B\}$) is given by
\begin{align}&
\cI_{\xa\beta}
= 
\frac{2\uppi g_{\xa\beta}^2}{\hbar}(1+\updelta_{\xa\beta})
\notag\\&\hphantom{\cI_{\xa\beta}{}={}}\times
\int\!\!\!\frac{\dd[3]{p_2}}{(2\uppi\hbar)^3}
\int\!\!\!\frac{\dd[3]{p_3}}{(2\uppi\hbar)^3}
\int\!\!\!\dd[3]{p_4}
\updelta_{\bp}(1234)
\updelta_E^{\xa\beta}(1234)
\notag\\&\hphantom{\cI_{\xa\beta}{}={}}\times
\Bigl\{
\bqty\big{1+\eta_\xa f_\xa(1)}\bqty\big{1+\eta_\beta f_\beta(2)}
f_\beta(3)f_\xa(4)
\notag\\&\hphantom{\cI_{\xa\beta}{}={}\times\Bigl\{}
-
f_\xa(1)f_\beta(2)
\bqty\big{1+\eta_\beta f_\beta(3)}\bqty\big{1+\eta_\xa f_\xa(4)}
\Bigr\}
\;,
\label{eq:collision_integral}
\end{align}
where we have used the abbreviated notation $f_\xa(i)=f_\xa(\br,\bp_i,t)\,(i=1,2,3,4)$ and taken $\eta_\B=+1$ and $\eta_\F=-1$, depending on the quantum
statistics~\cite{kadanoff_1962_WBI_quantum}.
On the right-hand side
of Eq.~\eqref{eq:collision_integral},
we have used the following notation:
\begin{subequations}
\begin{align}&
\updelta_{\bp}(1234)
=  
\updelta({\bp}_1+{\bp}_2-{\bp}_3-{\bp}_4)
\;,\\& 
\updelta_E^{\xa\beta}(1234)
= 
\updelta\biglb(\veps_\xa(\bp_1)+\veps_\beta(\bp_2)-\veps_\beta(\bp_3)-\veps_\xa(\bp_4)\bigrb)
\;.
\end{align}
\end{subequations}
These $\updelta$ functions reflect the fact that both momentum and energy are conserved in the binary collisions.
The equilibrium solution of the Boltzmann equation is given by the Bose--Einstein distribution function $f^0_\B=f^0_\B(\br,\bp)$ or Fermi--Dirac distribution function $f^0_\F=f^0_\F(\br,\bp)$:
\begin{align}&
f_\xa^0(\br,\bp)
=\frac1{\exp[\beta(\veps^0_\xa-\mu_\xa)]-\eta_\xa}
\;,
\label{eq:distribution_function_in_equilibrium}
\end{align}
where $\mu_\xa$ is the chemical potential, $\beta=1/(\kB T)$ is the inverse temperature, and $\veps^0_\xa=\veps^0_\xa(\br,\bp)$ is given by
Eqs.~\eqref{eq:single-particle_energy}
and~\eqref{eq:potential_energy},
with $n_\xa$ replaced with the equilibrium values $n^0_\xa=n^0_\xa(\br)$, which is self-consistently obtained using 
Eq.~\eqref{eq:distribution_function_in_equilibrium}
in Eq.~\eqref{eq:number_density}.

\subsection{Linearization}
To study dipole oscillation, we introduce the deviation of the distribution function from the equilibrium state, given by $\delta f_\xa=f_\xa-f_\xa^0$.
By assuming that our system is near equilibrium, we expand the Boltzmann
equation~\eqref{eq:Boltzmann_equation}
to first order in $\delta f_\xa$.
From this linearized version
of Eq.~\eqref{eq:Boltzmann_equation},
we can derive the equation of motion for the average value of an arbitrary physical quantity $\chi=\chi(\br,\bp)$ as
\begin{align}&
\dv{\delta\ev{\chi}_\xa}{t}
-\frac1{m_\xa}\delta\ev{\bp\cdot\pdv{\chi}{\br}}_\xa
+\delta\ev{\pdv{U_\xa^0}{\br}\cdot\pdv{\chi}{\bp}}_\xa
\notag\\&\hphantom{\dv{\delta\ev{\chi}_\xa}{t}}
-\ev{
\chi
\pdv{f^0_\xa}{\veps^0_\xa}\frac{\bp}{m_\xa}
\cdot
\pdv{\delta U_\xa}{\br}
}
=\ev{\chi\cI_\xa}
\;,
\label{eq:equation_of_motion_for_the_average}
\end{align}
where we define the following moments:
\begin{subequations}
\begin{align}&
\delta\ev{\chi}_\xa
=\frac1{N_\xa}\iint\!\frac{\dd[3]{r}\dd[3]{p}}{(2\uppi\hbar)^3}
\chi(\br,\bp)\delta f_\xa(\br,\bp,t)
\;,\\&
\ev{\chi A_\xa}
=\frac1{N_\xa}\iint\!\frac{\dd[3]{r}\dd[3]{p}}{(2\uppi\hbar)^3}
\chi(\br,\bp)A_\xa(\br,\bp,t)
\;,
\end{align}
\label{eq:moment}%
\end{subequations}
with the total number of particles $N_\xa$ and the arbitrary function $A_\xa=A_\xa(\br,\bp,t)$.
The fluctuation of the potential energy $\delta U_\xa$
in Eq.~\eqref{eq:equation_of_motion_for_the_average}
is given by $\delta U_\B=2g_{\BB}\delta n_\B+g_\BF\delta n_\F$ and $\delta U_\F=g_\BF\delta n_\B$, where the deviation of the number density $\delta n_\xa$ is also defined by $\delta n_\xa=n_\xa-n^0_\xa$.

\subsection{Moment equation}
Dipole oscillation is described by the displacement of the centers of mass of both bosonic and fermionic clouds.
Assuming dipole oscillation in the $z$ direction, we take $\chi=z,\,p_z$
in Eq.~\eqref{eq:equation_of_motion_for_the_average}.
We then obtain a coupled set of equations of motion for $\delta\ev{z}_\xa$ and $\delta\ev{v_z}_\xa=\delta\ev{p_z}_\xa/m_\xa$.
However, the resulting moment equations are generally not closed because of both the mean-field and collision terms.
To truncate these terms, we introduce the following ansatz:
\begin{align}&
\delta f_\xa
=\pdv{f^0_\xa}{\veps^0_\xa}
\pqty\bigg{a_\xa+\bb_\xa\cdot\bp+c_\xa\frac{p^2}{2m_\xa}}
\;,
\label{eq:ansatz_in_dipole_mode}
\end{align}
where we assume that $\bb_\xa(t)=b_\xa(t)\bhz$, and $b_\xa(t)$ does
not depend on the position.
We can relate $\bb_\xa$ to the velocity field by $\bv_\xa\equiv\delta\ev{v_z}_\xa\bhz=-\bb_\xa/3$.

Using the ansatz of
Eq.~\eqref{eq:ansatz_in_dipole_mode},
we obtain the following closed set of coupled moment equations 
for $\{\xa,\beta\}=\{\B,\F\}$ or $\{\F,\B\}$:
\begin{subequations}
\begin{align}&
\dv{\delta\ev{z}_\xa}{t}-\delta\ev{v_z}_\xa
=0
\label{eq:dipole_moment_eq_a}
\;,\\&
\dv{\delta\ev{v_z}_\xa}{t}
+\pqty{
\zomega_\xa^2
-\frac{\Delta}{M_\xa}
}\delta\ev{z}_\xa
\notag\\&\hphantom{\dv{\delta\ev{v_z}_\xa}{t}}
+\frac{\Delta}{M_\xa}\delta\ev{z}_\beta
=-\frac{M_+}{\tau M_\xa}
\pqty\big{\delta\ev{v_z}_\xa-\delta\ev{v_z}_\beta}
\;,
\label{eq:dipole_moment_eq_b}
\end{align}
\label{eq:dipole_moment_eq}%
\end{subequations}
where $1/{M_\pm}=1/{M_\B}\pm1/{M_\F}$ with $M_\xa=m_\xa N_\xa$, and $\Delta$ is the mean-field contribution, which depends on the equilibrium density profiles through 
\begin{align}& 
\Delta
=g_\BF\int\!\!\!\dd[3]{r}
\pdv{n^0_\B}{z}\pdv{n^0_\F}{z}
\;.
\label{eq:mean_field_contribution_in_dipole_mode}
\end{align}
Regarding the collision terms, the contribution $\ev{z\cI_\xa}$ vanishes, because the collisions conserve the number of particles of each component.
We also find that $\ev{p_z\cI_\BB}$ vanishes, because intraspecies (Bose--Bose) collisions conserve the momentum of bosons.
In contrast, interspecies (Bose--Fermi) collisions do not conserve the momentum of each component independently but conserve the total momentum.
We thus have nonzero contributions from $\ev{p_z\cI_\BF}$ and $\ev{p_z\cI_\FB}$.
Within our truncation scheme, the collisional contributions are given by 
\begin{subequations}
\begin{align}&
\ev{p_z\cI_\BF}
=-\frac{M_+}{\tau N_\B}
\pqty\big{\delta\ev{v_z}_\B-\delta\ev{v_z}_\F}
\;,\\&
\ev{p_z\cI_\FB}
=-\frac{M_+}{\tau N_\F}
\pqty\big{\delta\ev{v_z}_\F-\delta\ev{v_z}_\B}
\;,
\end{align}
\end{subequations}
where the characteristic relaxation time $\tau$ is given by 
\begin{align}&
\frac1{\tau}
=\frac{3\uppi\beta g_\BF^2}{\hbar M_+}
\int\!\!\!\dd[3]{r}
\int\!\!\!\frac{\dd[3]{p_1}}{(2\uppi\hbar)^3}
\int\!\!\!\frac{\dd[3]{p_2}}{(2\uppi\hbar)^3}
\int\!\!\!\frac{\dd[3]{p_3}}{(2\uppi\hbar)^3}
\int\!\!\!\dd[3]{p_4}
\notag\\&\hphantom{\frac1{\tau}{}={}\frac{3\uppi\beta g_\BF^2}{\hbar M_+}}\times 
\updelta_{\bp}(1234)
\updelta_E^\BF(1234)\pqty{p_{1z}-p_{4z}}^2
\notag\\&\hphantom{\frac1{\tau}{}={}\frac{3\uppi\beta g_\BF^2}{\hbar M_+}}\times
\bqty\big{1+f^0_\B(1)}\bqty\big{1-f^0_\F(2)}
f^0_\F(3)f^0_\B(4)
\; .
\label{eq:collision_rate_in_dipole_mode}
\end{align} 
Equation~\eqref{eq:dipole_moment_eq}
can be reduced into the following form:
\begin{align}&
\dv{}{t}{\bm u}
= 
A 
\bu 
\;,
\label{eq:dipole_moment_eq_in_matrix_form}
\end{align}
where 
$\bu=\bqty{\delta\ev{z}_\B,\delta\ev{z}_\F,\delta\ev{v_z}_\B,\delta\ev{v_z}_\F}^\mathrm{T}$, and 
\begin{align}&
A=
\pmqty{%
0&0&1&0
\\
0&0&0&1
\\
\frac{\Delta}{M_\B}-\zomega_\B^2&-\frac{\Delta}{M_\B}&-\frac{1}{\tau}\frac{M_+}{M_\B}& \frac{1}{\tau}\frac{M_+}{M_\B}
\\[1ex]
-\frac{\Delta}{M_\F}&\frac{\Delta}{M_\F}-\zomega_\F^2& \frac{1}{\tau}\frac{M_+}{M_\F}&-\frac{1}{\tau}\frac{M_+}{M_\F}
}
\;.
\end{align}
By considering the normal-mode solution $\bu\propto\ue^{-\ui\omega t}$
of Eq.~\eqref{eq:dipole_moment_eq_in_matrix_form}, 
we obtain the following quartic equation of $\omega$ for nontrivial solutions:
\begin{align}&
\pqty{\omega^2-\zomega_\B^2+\frac{\Delta}{M_\B}}
\pqty{\omega^2-\zomega_\F^2+\frac{\Delta}{M_\F}}
-\frac{\Delta^2}{M_\B M_\F}
\notag\\&
+\frac{\ui\omega M_+}{\tau}
\pqty{
\frac{\omega^2-\zomega_\F^2}{M_\B}
+\frac{\omega^2-\zomega_\B^2}{M_\F}
}
=0
\;.
\label{eq:secular_equation_in_dipole_mode_without_anisotropy}
\end{align}

We note that the coupled set of
equations~\eqref{eq:dipole_moment_eq}
is obtained from the Boltzmann
equation~\eqref{eq:Boltzmann_equation},
and the mean-field contribution $\Delta$ and the relaxation time $\tau$ are microscopically introduced, which include the quantum statistical effects.
In particular, the term $\Delta$ originates from the mean-field potential $g_\BF n_{\F(\B)}$
in Eq.~\eqref{eq:potential_energy}.
In the absence of $\Delta$, 
Eq.~\eqref{eq:dipole_moment_eq} is consistent with a classical model for two harmonic oscillators coupled through phenomenologically included collisional damping terms~\cite{ferrari_2002_PRL_89_collisional,ferlaino_2003_JOBQSO_5_dipolar}.

We find
from Eqs.~\eqref{eq:dipole_moment_eq}
and~\eqref{eq:secular_equation_in_dipole_mode_without_anisotropy} 
that only the interspecies interaction has an explicit effect on the dipole modes of a two-component mixture through the mean-field
contribution~\eqref{eq:mean_field_contribution_in_dipole_mode}
and the relaxation
time~\eqref{eq:collision_rate_in_dipole_mode}.
The effect of the intraspecies interaction is only implicitly included in the equilibrium property through the effective potential for bosons in
Eq.~\eqref{eq:potential_energy_for_a_boson}.
This implies that the formulation of the dipole modes in this paper can be easily extended to other types of two-component mixtures, such as Bose--Bose and Fermi--Fermi mixtures with $s$-wave scattering interactions.
In general, the moment equations for dipole modes are given in the form of
\eqref{eq:dipole_moment_eq}
and~\eqref{eq:secular_equation_in_dipole_mode_without_anisotropy},
where a mean-field contribution $\Delta$ and a relaxation time $\tau$ include the effect of interspecies interactions.
The temperature dependence of the dipole modes may be different, depending on the quantum statistics of the two-component mixture, reflecting the different temperature dependences of $\Delta$ and $\tau$.
We note that the importance of the interspecies interaction in the collective mode was also found in the presence of BEC by using the sum-rule approach 
at $T=0$~\cite{miyakawa_2000_PRA_62_sum}.
In Ref.~\cite{miyakawa_2000_PRA_62_sum},
the effect of the interspecies interaction is described
by Eq.~\eqref{eq:collision_integral},
which is similar
to our Eq.~\eqref{eq:mean_field_contribution_in_dipole_mode},
but uses the condensate density instead
of Eq.~\eqref{eq:number_density}
for $n_\B$ in this paper.

\section{Results}
\label{sec:result}
In this section, we discuss the results obtained from the moment
equation~\eqref{eq:dipole_moment_eq_in_matrix_form}
for the dipole modes.
To extract analytic expressions of $\omega=\Omega-\ui\Gamma$ that determine both the frequency $\Omega$ and damping rate $\Gamma$, we first focus on two limiting cases: the collisionless regime
(Sec.~\ref{subsec:collisionless_limit})
and the hydrodynamic regime
(Sec.~\ref{subsec:hydrodynamic_limit}).
We also determine how the collisionless dipole modes turn into hydrodynamic dipole modes 
as a function of the relaxation time in the absence of $\Delta$
(Sec.~\ref{subsec:behavior_between_both_regimes}), 
and as a function of the temperature where the mean-field contribution $\Delta$ is fully included
(Sec.~\ref{subsec:temperature_dependence}).
Finally, we discuss the connection with experiments on dipole modes
(Sec.~\ref{discussions_with_experiments}).

\subsection{Collisionless limit}
\label{subsec:collisionless_limit}
In the collisionless limit $\zomega_{\B,\F}\tau\gg1$,
from Eq.~\eqref{eq:secular_equation_in_dipole_mode_without_anisotropy} 
we obtain two types of oscillating modes $(\Omega_+,\Gamma_+)$ and 
$(\Omega_-,\Gamma_-)$, 
where the frequency $\Omega_\pm$ and the damping rate $\Gamma_\pm$ are given by
\begin{subequations}
\begin{align}&
\Omega_\pm^2= 
\omega_+^2-\frac{\Delta}{2M_+}
\pm
\sqrt{%
\omega_-^2
\pqty{\omega_-^2-\frac{\Delta}{M_-}}
+\frac{\Delta^2}{4M_+^2}
}
\;, 
\label{eq:eigenfrequency_in_collisionless_limit_in_dipole_mode}
\\&
\Gamma_\pm 
=  
\frac1{4\tau}
\bqty{%
1\pm 
\frac1{\Omega_+^2-\Omega_-^2}
\pqty{\frac{2M_+}{M_-}\omega_-^2
-\frac{\Delta}{M_+}}
}
\;,
\label{eq:damping_rate_in_collisionless_limit_in_dipole_mode}
\end{align}
\label{eq:eigenfrequency_and_damping_rate_in_collisionless_limit_in_dipole_mode}%
\end{subequations}
where $\omega_\pm^2=\pqty\big{\zomega_\B^2\pm\zomega_\F^2}/{2}$.
In the special case of $\zomega_\B=\zomega_\F$, we can easily see that $\Omega_+=\zomega_\B=\zomega_\F$ and $\Gamma_+=0$, which corresponds to the well-known Kohn 
mode~\cite{kohn_1961_PR_123_cyclotron}
(the undamped dipole oscillation independent of interactions, temperature, and quantum statistics).

We now discuss the eigenvectors in the collisionless limit in the case where $\zomega_\B\neq\zomega_\F$.
The physics of the eigenmodes in the collisionless regime in this case are more clearly shown for the case of $\Delta=0$, where the frequency and damping rate are given by
\begin{subequations}
\begin{align}&
\pqty\big{\Omega_+^2,\;\Omega_-^2}=
\pqty\big{\zomega_\B^2,\;\zomega_\F^2}
\;,\\&
\pqty\big{\Gamma_+,\;\Gamma_-}=
\pqty{\frac{M_+}{2\tau M_\B},\;\frac{M_+}{2\tau M_\F}}
\;.
\end{align}
\label{eq:eigenvalue_in_collisionless_limit_in_dipole_mode_without_Delta}%
\end{subequations}
In this case, the frequencies of the two types of oscillating modes coincide with the harmonic trap frequencies.
Eigenvectors associated with these modes are given by
\begin{align}&
\bu^+\equiv
\pmqty{%
\delta\ev{z}_\B^+
\\
\delta\ev{z}_\F^+
\\
\delta\ev{v_z}_\B^+
\\
\delta\ev{v_z}_\F^+
}
= 
\pmqty{%
\pqty\big{1-\zomega_\F^2/\zomega_\B^2}/M_+
\\
\ui/(\zomega_\B\tau M_\F)
\\
-\ui\zomega_\B\pqty\big{1-\zomega_\F^2/\zomega_\B^2}/M_+
\\
1/(\tau M_\F)
}
\;,
\label{eq:eigenvector_in_collisionless_b_mode}
\\&
\bu^-\equiv
\pmqty{%
\delta\ev{z}^-_\B
\\
\delta\ev{z}^-_\F
\\
\delta\ev{v_z}^-_\B
\\
\delta\ev{v_z}^-_\F
}
= 
\pmqty{%
\ui/(\zomega_\F\tau M_\B)
\\
\pqty\big{1-\zomega_\B^2/\zomega_\F^2}/M_+
\\
1/(\tau M_\B)
\\
-\ui\zomega_\F\pqty\big{1-\zomega_\B^2/\zomega_\F^2}/M_+
}
\;, 
\label{eq:eigenvector_in_collisionless_f_mode}
\end{align}
regardless of normalization.
In the $\Omega_+$ mode, the relation $\delta\ev{z}_\F^+/\delta\ev{z}_\B^+=\delta\ev{v_z}_\F^+/\delta\ev{v_z}_\B^+\propto\ui/\tau\to0$ applies, which means that the fermionic cloud shows negligibly small oscillations of the center of mass as well as the velocity field in the collisionless limit $\tau\to\infty$.
In this $\Omega_+$ mode, an oscillation is mainly due to the bosonic cloud with the harmonic trap frequency $\Omega_+=\zomega_\B$.
In the $\Omega_-$ mode, in contrast, the fermionic cloud largely oscillates with the harmonic trap frequency $\Omega_-=\zomega_\F$, providing the relation $\delta\ev{z}^-_\B/\delta\ev{z}^-_\F=\delta\ev{v_z}^-_\B/\delta\ev{v_z}^-_\F\propto\ui/\tau\to0$.
In this limit ($\zomega_{\B,\F}\tau\gg1$ and $\Delta\to0$), the ratio of the fluctuation $\delta\ev{z}^\pm_\B/\delta\ev{z}^\pm_\F=\delta\ev{v_z}^\pm_\B/\delta\ev{v_z}^\pm_\F$ is a purely imaginary number, and hence the phase difference between the two interspecies components is $\uppi/2$.
This anomalous phase difference $\uppi/2$ indicates that these modes are neither in-phase nor out-of-phase modes, which arises from the fact that the bosonic and fermionic clouds are coupled only through the collision term
[right-hand side of Eq.~\eqref{eq:dipole_moment_eq_b}].

\subsection{Hydrodynamic limit}
\label{subsec:hydrodynamic_limit}
In the hydrodynamic limit $\zomega_{\B,\F}\tau\ll1$, 
we obtain an oscillating mode and two purely damped relaxation modes (fast and slow relaxing modes).
The eigenfrequencies of these modes are respectively given by 
\begin{align}&
\omega
=
\begin{dcases}
\Omega_\mix-\ui\Gamma_\mix
\;,\\ 
\hphantom{\Omega_\mix}
-\ui\Gamma_\fast
\;,\\
\hphantom{\Omega_\mix}
-\ui\Gamma_\slow 
\;, 
\end{dcases}
\label{eq:eigenvalue_in_hydrodynamic_limit_in_dipole_mode}
\end{align}
where 
\begin{subequations}
\begin{align}&
\Omega_\mix^2
=\frac{M_\B\zomega_\B^2+M_\F\zomega_\F^2}{M_\B+M_\F}
\label{eq:frequency_of_in-phase_mode}
\;,\\&
\Gamma_\mix
=\tau\frac{M_+\pqty\big{\zomega_\B^2-\zomega_\F^2}^2}{2\pqty\big{M_\B\zomega_\B^2+M_\F\zomega_\F^2}}
\;,\\&
\Gamma_\fast
=\frac1{\tau}
\;,\\&
\Gamma_\slow
=\tau\pqty{\frac{\zomega_\B^2\zomega_\F^2}{\Omega_\mix^2}-\frac{\Delta}{{M_+}}}
\;.
\end{align}
\end{subequations}

\begin{figure}
\centering
\includegraphics[scale=.62]{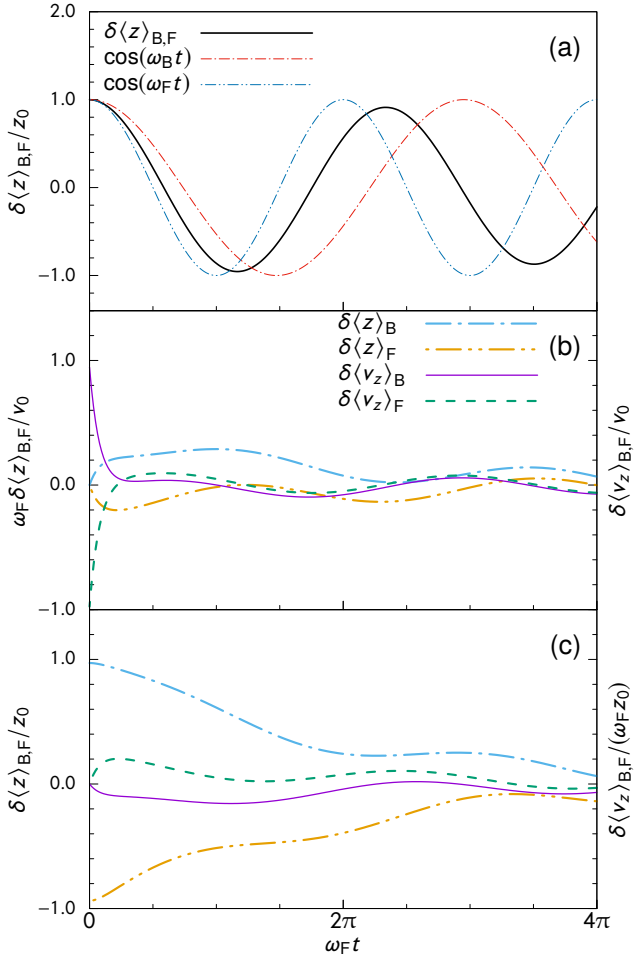}
\caption{
Time evolution of the center-of-mass positions and velocity fields, assuming a Bose--Fermi mixture of $^{87}\mathrm{Rb}$ and $^{40}\mathrm{K}$, where the harmonic trap potentials for both components are the same, giving the relation $m_\B\zomega_\B^2=m_\F\zomega_\F^2$, and $m_\B/m_\F=87/40$.
All the panels show the case where $\Delta$ is absent and $\zomega_\F\tau=0.25$.
(a)~Characteristic oscillation of the in-phase mode with frequency $\Omega_\mix$ generated from the initial condition given by $\delta\ev{z}_{\B,\F}=z_0$ and $\delta\ev{v_z}_{\B,\F}=0$.
For comparison, we also plot $\cos(\zomega_\B t)$ and $\cos(\zomega_\F t)$ with the bare trap frequencies $\zomega_{\B,\F}$.
(b)~Characteristic oscillation of the fast relaxing mode generated from the initial condition given by $\delta\ev{z}_{\B,\F}=0$ and $\delta\ev{v_z}_\B=-\delta\ev{v_z}_\F=v_0$.
In~(a) and~(b), we take $N_\B/N_\F=m_\F/m_\B$, which provides $M_\B=M_\F$.
(c)~Characteristic oscillation of the slow relaxing mode generated from the initial condition given by $\delta\ev{z}_\B=-\delta\ev{z}_\F=z_0$ and $\delta\ev{v_z}_{\B,\F}=0$.
Here, we take $N_\B=N_\F$, which provides $M_\B\zomega_\B^2=M_\F\zomega_\F^2$.
}
\label{fig:time_evolution}
\end{figure}

The mode with $\omega=\Omega_\mix-\ui\Gamma_\mix$ represents the in-phase oscillation of the two components.
If we neglect $\Delta$, the eigenvector of the in-phase mode is simply given by 
\begin{align}&
\bu^\mix\equiv 
\pmqty{%
\delta\ev{z}_\B^\mix
\\[2pt]
\delta\ev{z}_\F^\mix
\\[2pt]
\delta\ev{v_z}_\B^\mix
\\[2pt]
\delta\ev{v_z}_\F^\mix
}
= 
\pmqty{%
1
\\[2pt]
1
\\[2pt]
-\ui\Omega_\mix
\\[2pt]
-\ui\Omega_\mix
}
\;, 
\label{eq:eigenvector_in_hydrodynamic_oscillationg_modes}
\end{align}
regardless of normalization.
We immediately find the relation $\delta\ev{z}_\B/\delta\ev{z}_\F=\delta\ev{v_z}_\B/\delta\ev{v_z}_\F=1\,(>0)$, which indicates that in this in-phase mode, fluctuations of both the bosonic and fermionic clouds are equivalent: $\delta\ev{z}_\B=\delta\ev{z}_\F$ and $\delta\ev{v_z}_\B=\delta\ev{v_z}_\F$.
In the hydrodynamic regime, Bose--Fermi collisions bring the mixture gas to local equilibrium so fast that one component is dragged and accompanied by the other component.
We note that even for $\zomega_\B\neq\zomega_\F$, the two clouds oscillate at the same frequency $\Omega_\mix$, as if they were a single cloud
[Fig.~\ref{fig:time_evolution}(a)].
In the case of equal trap frequencies, i.e., $\zomega_\B=\zomega_\F$, the in-phase mode reduces to the undamped Kohn mode with $\Omega_\mix=\zomega_\B=\zomega_\F$ and $\Gamma_\mix=0$, as in the collisionless limit.

One of the two purely damped relaxation modes is the fast relaxing mode with a damping rate of $\Gamma_\fast$ ($\propto1/\tau$).
The other is the slow relaxing mode with a damping rate of $\Gamma_\slow$ ($\propto\tau$).
Both the purely damped relaxation modes involve out-of-phase motion of the two components.
Indeed, in the case of $\Delta=0$, the eigenvectors of the fast and slow relaxing modes are respectively given by 
\begin{align}&
\bu^\fast\equiv
\pmqty{%
\delta\ev{z}^\fast_\B
\\[2pt]
\delta\ev{z}^\fast_\F
\\[2pt]
\delta\ev{v_z}^\fast_\B
\\[2pt]
\delta\ev{v_z}^\fast_\F
}
= 
\pmqty{%
\tau/M_\B
\\[2pt]
-\tau/M_\F
\\[2pt]
-1/M_\B
\\[2pt]
1/M_\F
}
\;,
\label{eq:eigenvector_in_hydrodynamic_fast_relaxing_mode}
\\& 
\bu^\slow\equiv
\pmqty{%
\delta\ev{z}^\slow_\B
\\[2pt]
\delta\ev{z}^\slow_\F
\\[2pt]
\delta\ev{v_z}^\slow_\B
\\[2pt]
\delta\ev{v_z}^\slow_\F
}
= 
\pmqty{%
\zomega_\F^2/M_\B
\\[2pt]
-\zomega_\B^2/M_\F
\\[2pt]
-{\tau\zomega_\B^2\zomega_\F^4}/{(\Omega_\mix^2M_\B)}
\\[2pt]
{\tau\zomega_\B^4\zomega_\F^2}/{(\Omega_\mix^2M_\F)}
}
\;, 
\label{eq:eigenvector_in_hydrodynamic_slow_decay_mode}
\end{align}
regardless of normalization.
Both Eqs.~\eqref{eq:eigenvector_in_hydrodynamic_fast_relaxing_mode}
and~\eqref{eq:eigenvector_in_hydrodynamic_slow_decay_mode}
give the relation $\delta\ev{z}_\B/\delta\ev{z}_\F=\delta\ev{v_z}_\B/\delta\ev{v_z}_\F<0$, indicating the out-of-phase mode.

We can see distinct features of the two relaxation modes from the eigenvectors
in Eqs.~\eqref{eq:eigenvector_in_hydrodynamic_fast_relaxing_mode}
and~\eqref{eq:eigenvector_in_hydrodynamic_slow_decay_mode}.
From Eq.~\eqref{eq:eigenvector_in_hydrodynamic_fast_relaxing_mode}, 
the fast relaxing mode involves a large relative velocity and a small spatial deviation, since $\delta\ev{z}^\fast_{\B,\F}/\delta\ev{v_z}^\fast_{\B,\F}=-\tau\to0$.
In contrast,
from Eq.~\eqref{eq:eigenvector_in_hydrodynamic_slow_decay_mode},
the slow relaxation mode involves a small relative velocity and a large relative 
spatial deviation, since $\delta\ev{v_z}^\slow_{\B,\F}/\delta\ev{z}^\slow_{\B,\F}=-\tau\zomega_\B^2\zomega_\F^2/\Omega_\mix^2\to0$.

These two distinct relaxation modes can be selectively excited by choosing an appropriate initial state.
If initially two atomic clouds substantially overlap at the center of the trap with a large relative momentum, the system approaches the static equilibrium quickly.
This type of fast relaxation mode is also present in a uniform mixture as a diffusion mode.
In contrast, if the two centers of mass are sufficiently separated in space with a small relative momentum as an initial state,
then the two atomic clouds are gradually mixed through a long relaxation process.
We note that this type of slow relaxation mode is unique to a trapped system, since the damping rate of this mode vanishes in the limit where the trap frequencies are zeros.
These facts will be clearly seen in the discussion in the following  paragraphs
[see Figs.~\ref{fig:time_evolution}(b) and~\ref{fig:time_evolution}(c)].

To be more explicit, we express a general dipole motion in the hydrodynamic limit as a linear combination of the in-phase oscillating mode and the two purely damped modes.
Using
Eqs.~\eqref{eq:eigenvalue_in_hydrodynamic_limit_in_dipole_mode},
\eqref{eq:eigenvector_in_hydrodynamic_oscillationg_modes},
\eqref{eq:eigenvector_in_hydrodynamic_fast_relaxing_mode},
and~\eqref{eq:eigenvector_in_hydrodynamic_slow_decay_mode}
in the case of $\Delta=0$, we have the expression 
\begin{align}&
\bu 
=\Re\bqty{C_\mix
\bu^\mix
\exp(-\ui\Omega_\mix t)}\exp(-\Gamma_\mix t)
\notag\\&\hphantom{\bu{}={}}
+C_\fast
\bu^\fast 
\exp(-\Gamma_\fast t)
+C_\slow
\bu^\slow 
\exp(-\Gamma_\slow t)
\;, 
\label{eq:liner_combination_of_dipole_mode_in_hydrodynamic_regime}
\end{align}
where $C_\mix$, $C_\fast$, and $C_\slow$ are coefficients determined by the initial condition.

For example, if we take the initial condition where two centers of mass lie on an equal position without initial velocities, i.e., $\delta\ev{z}_{\B,\F}=z_0$ and $\delta\ev{v_z}_{\B,\F}=0$, the weights of the two relaxation modes disappear as $C_\fast=C_\slow=0$, but the weight of the in-phase mode remains as $C_\mix=z_0$, resulting in
\begin{align}&
\bu=z_0\Re
\bqty{
\bu^\mix\exp(-\ui\Omega_\mix t)
}
\exp(-\Gamma_\mix t)
\;.
\label{eq:in_phase_oscillation_with_an_initial_state}
\end{align}
Here, both the bosonic and fermionic clouds oscillate at the same frequency $\Omega_\mix$ despite the different trap frequencies $\zomega_\B\neq\zomega_\F$, as shown
in Fig.~\ref{fig:time_evolution}(a).

In the initial condition where two atomic clouds at the center of the trap are kicked in opposite directions, i.e., $\delta\ev{v_z}_\B=-\delta\ev{v_z}_\F=v_0$ and $\delta\ev{z}_{\B,\F}=0$, the coefficients are approximately given to the first order in $\tau$ as
\begin{subequations}
\begin{align}&
C_\mix=\frac{2\tau v_0M_+}{\Omega_\mix^2}\frac{\zomega_\B^2-\zomega_\F^2}{M_\B+M_\F}
+\ui\frac{v_0}{\Omega_\mix}\frac{M_\B-M_\F}{M_\B+M_\F}
\;,\\&
C_\fast=-2v_0M_+
\;,\\&
C_\slow=\frac{2\tau v_0M_+}{\Omega_\mix^2}
\;.
\end{align}
\label{eq:initial_state_to_tominate_fast_relaxing_mode}%
\end{subequations}

In the hydrodynamic limit $\tau\to0$, the weight of the slow relaxing mode $C_\slow$ is negligibly small.
In particular, in the case of $\zomega_\B=\zomega_\F$ and $M_\B=M_\F$, we have $C_\mix=0$.
Consequently, only the fast relaxing mode is excited.
Figure~\ref{fig:time_evolution}(b)
shows the time evolution
of Eq.~\eqref{eq:liner_combination_of_dipole_mode_in_hydrodynamic_regime}
with this initial condition.
We observe a strong damping of the velocity fields, 
although the small-amplitude oscillation emerges because of the contribution from the in-phase mode.

In the initial condition where two atomic clouds are spatially separated without initial velocities, i.e., $\delta\ev{z}_\B=-\delta\ev{z}_\F=z_0$ and $\delta\ev{v_z}_{\B,\F}=0$, the coefficients are approximately given to first order in $\tau$ as
\begin{subequations}
\begin{align}&
C_\mix=z_0\frac{M_\B\zomega_\B^2-M_\F\zomega_\F^2}{M_\B\zomega_\B^2+M_\F\zomega_\F^2}
-\ui\tau\frac{2z_0M_+\zomega_\B^2\zomega_\F^2}{\Omega_\mix^5}\frac{\zomega_\B^2-\zomega_\F^2}{M_\B+M_\F}
\;,\\&
C_\fast=-2\tau z_0M_+\frac{\zomega_\B^2\zomega_\F^2}{\Omega_\mix^2}
\;,\\&
C_\slow=\frac{2z_0M_+}{\Omega_\mix^2}
\;.
\end{align}
\label{eq:initial_state_to_tominate_slow_decay_mode}%
\end{subequations}

The weight of the fast relaxing mode $C_\fast$ is negligibly small in the hydrodynamic limit $\tau\to0$.
Again, in the case of $\zomega_\B=\zomega_\F$ and $M_\B=M_\F$, we have $C_\mix=0$.
In this case, only the slow relaxing mode is excited.
Figure~\ref{fig:time_evolution}(c)
shows the time evolution
of Eq.~\eqref{eq:liner_combination_of_dipole_mode_in_hydrodynamic_regime}
for this initial condition, which clearly exhibits that the characteristic feature of the slow relaxing mode emerges.

\subsection{Behavior in the transition between collisionless and hydrodynamic regimes}
\label{subsec:behavior_between_both_regimes}
\begin{figure*}
\includegraphics[scale=.62]{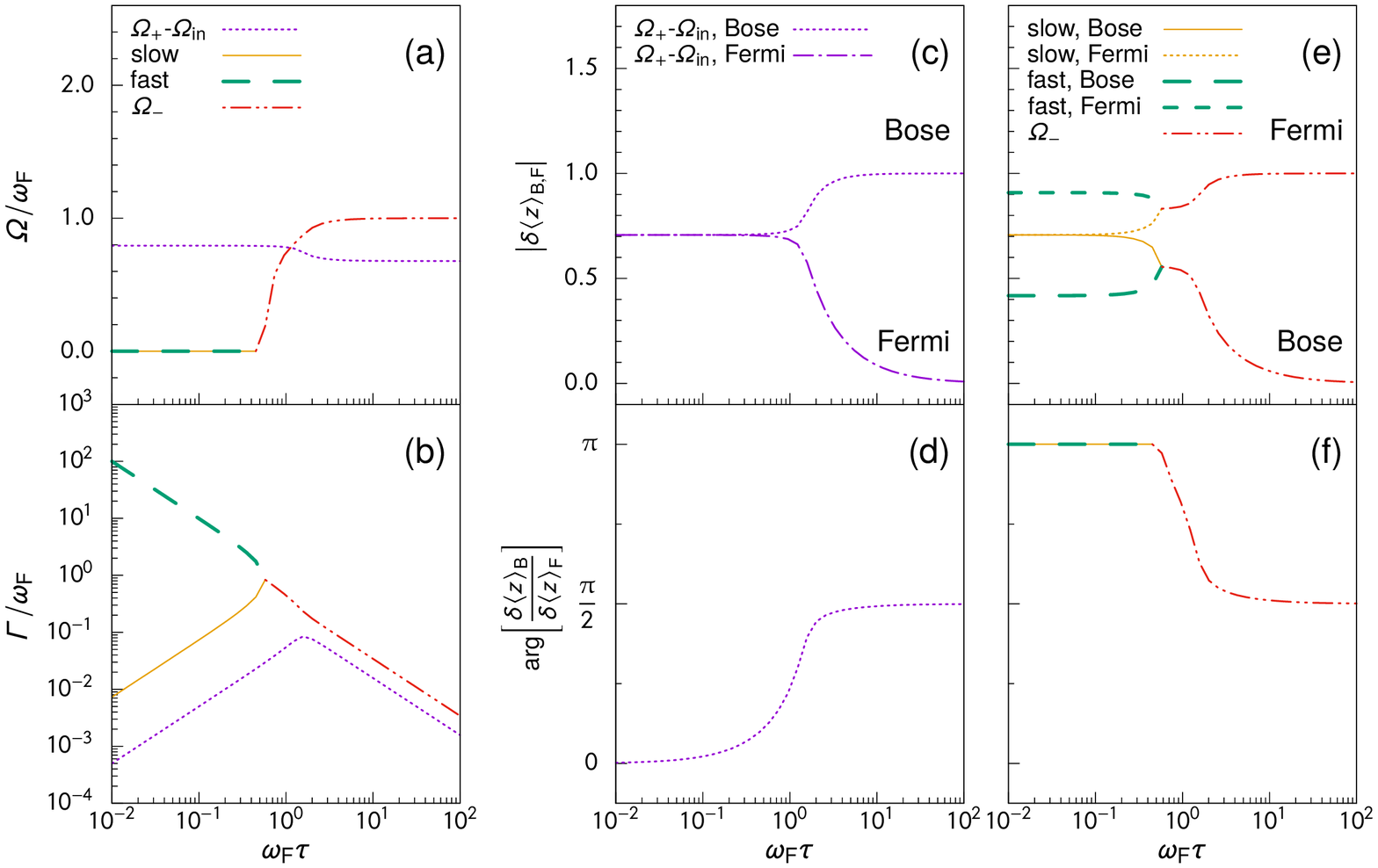}
\caption{
Dipole modes as a function of the relaxation time $\tau$ in the absence of $\Delta$.
(a)~Frequency $\Omega$ and (b)~damping rate $\Gamma$.
(c)~Mode amplitude and (d)~relative phase between $\delta\ev{z}_\B$ and $\delta\ev{z}_\F$ for the $\Omega_+$ (or $\Omega_\mix$) mode, given by the eigenvector
of Eq.~\eqref{eq:dipole_moment_eq_in_matrix_form}.
The mode amplitudes and the relative phase for the $\Omega_-$ (or relaxation) mode are shown in panels~(e) and~(f); the mode amplitudes are normalized as $\abs*{\delta\ev{z}_\B}^2+\abs*{\delta\ev{z}_\F}^2=1$.
The results of the velocity fields $\delta\ev{v_z}_{\B,\F}$ are the same as for $\delta\ev{z}_{\B,\F}$.
We assume a mixture of $^{87}\mathrm{Rb}$ and $^{40}\mathrm{K}$ with $N_\B=N_\F$ and $m_\B\zomega_\B^2 = m_\F \zomega_\F^2$.
}
\label{fig:000000_di_eig_val_vec_01}
\end{figure*}
\begin{figure*}
\centering
\includegraphics[scale=.62]{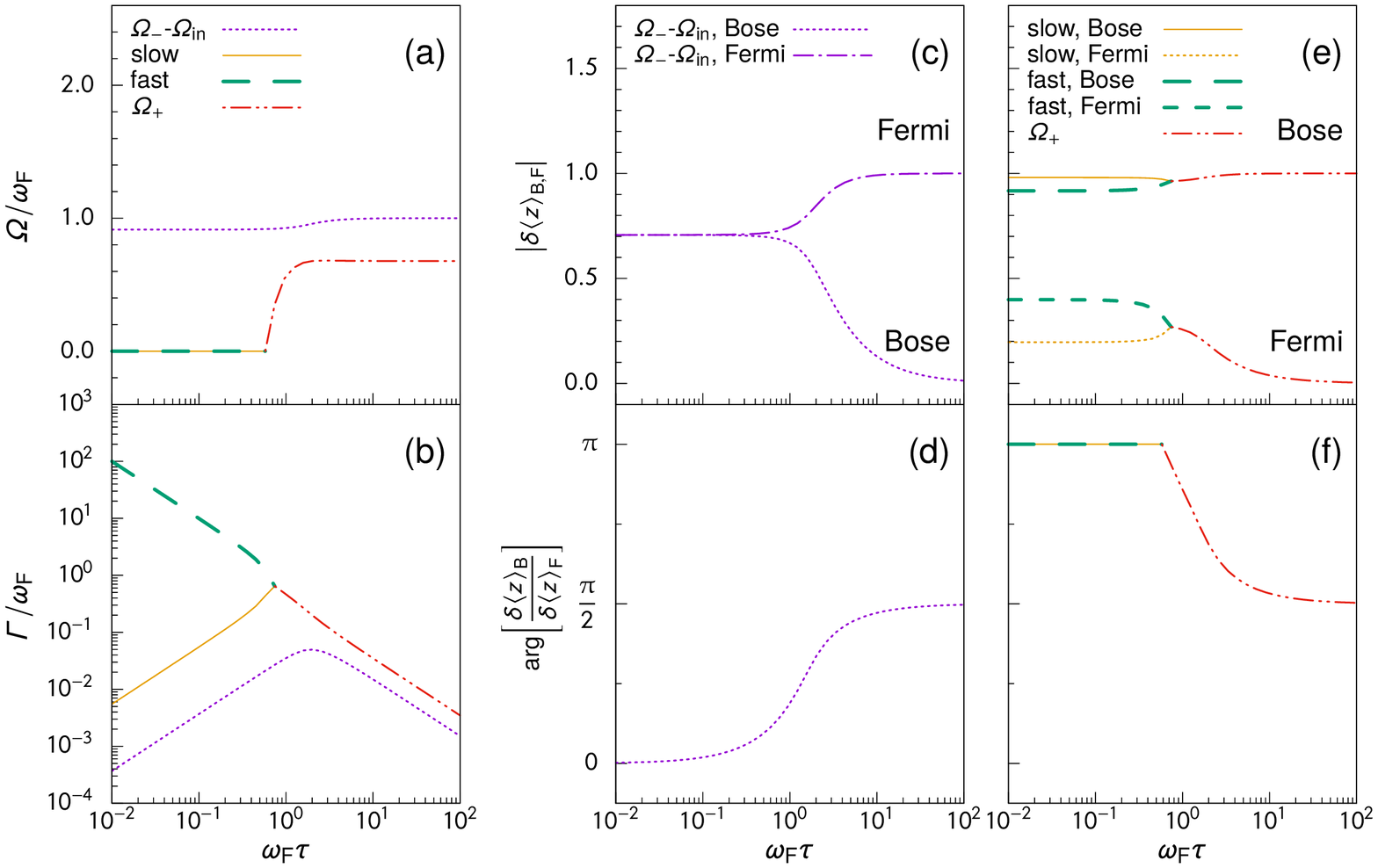}
\caption{
As for Fig.~\ref{fig:000000_di_eig_val_vec_01}
but for $N_\B/N_\F=0.2$.
The $\Omega_-$ mode smoothly connects to the in-phase mode in the hydrodynamic regime, in contrast to the case
in Fig.~\ref{fig:000000_di_eig_val_vec_01}.
}
\label{fig:000010_di_eig_val_vec_01}
\end{figure*}
Figures~\ref{fig:000000_di_eig_val_vec_01}(a) and~\ref{fig:000000_di_eig_val_vec_01}(b)
show the relaxation time dependencies of the frequency and the damping rate in the absence of the mean-field contribution ($\Delta=0$), respectively.
The frequency of the $\Omega_+$ mode in the collisionless regime shows a smooth crossover to the in-phase mode with decreasing relaxation time $\tau$
[Fig.~\ref{fig:000000_di_eig_val_vec_01}(a)].
In the crossover regime, the lifetime of this mode becomes the shortest
[Fig.~\ref{fig:000000_di_eig_val_vec_01}(b)].
In contrast, the frequency of the $\Omega_-$ mode in the collisionless regime drops to zero at a certain value of the relaxation time
[Fig.~\ref{fig:000000_di_eig_val_vec_01}(a)].
After this transition, the damping rate of this mode shows a bifurcation, where both the fast and slow relaxing modes emerge
[Fig.~\ref{fig:000000_di_eig_val_vec_01}(b)].
These behaviors of the eigenvalues indicate that not all the dipole modes exhibit a smooth crossover, as in the case of the crossover between the zero sound mode and the first sound mode.

For the $\Omega_+$ mode in the collisionless regime, the oscillation of the center of mass is dominated by the bosonic cloud
[Fig.~\ref{fig:000000_di_eig_val_vec_01}(c)],
where the phase difference is $\uppi/2$
[Fig.~\ref{fig:000000_di_eig_val_vec_01}(d)].
With decreasing relaxation time, the in-phase mode emerges with frequency $\Omega_\mix$, where both the bosonic and fermionic clouds oscillate
[Fig.~\ref{fig:000000_di_eig_val_vec_01}(c)].
The amplitudes of this mode are the same for both the bosonic and fermionic clouds in the hydrodynamic regime
[Fig.~\ref{fig:000000_di_eig_val_vec_01}(c)],
where the phase difference approaches zero 
[Fig.~\ref{fig:000000_di_eig_val_vec_01}(d)].

For the $\Omega_-$ mode in the collisionless regime, the oscillation of the center of mass is dominated by the fermionic cloud
[Fig.~\ref{fig:000000_di_eig_val_vec_01}(e)]
and the phase difference is also $\uppi/2$
[Fig.~\ref{fig:000000_di_eig_val_vec_01}(f)].
With decreasing relaxation time, there occurs a transition from the $\Omega_-$ mode to the two purely damped modes, where the amplitudes of both the bosonic and fermionic clouds bifurcate 
[Fig.~\ref{fig:000000_di_eig_val_vec_01}(e)].
In the hydrodynamic regime, the phase difference of these purely damped modes approaches $\uppi$, which clearly indicates out-of-phase motion
[Fig.~\ref{fig:000000_di_eig_val_vec_01}(f)].

While Fig.~\ref{fig:000000_di_eig_val_vec_01}
shows that the bosonic cloud oscillates in the whole regime, 
the opposite situation can be produced, as shown
in Fig.~\ref{fig:000010_di_eig_val_vec_01},
where the fermionic cloud oscillates in the whole regime.
Figure~\ref{fig:000010_di_eig_val_vec_01}
is the same as
Fig.~\ref{fig:000000_di_eig_val_vec_01},
but for a different number ratio $N_\B/N_\F=0.2$
(compared to $N_\B=N_\F$
for Fig.~\ref{fig:000000_di_eig_val_vec_01}).
In the collisionless regime, the $\Omega_+$ mode and $\Omega_-$ mode are dominated by the bosonic cloud
[Fig.~\ref{fig:000010_di_eig_val_vec_01}(e)]
and the fermionic cloud
[Fig.~\ref{fig:000010_di_eig_val_vec_01}(c)],
respectively.
However, as opposed to the case shown
in Fig.~\ref{fig:000000_di_eig_val_vec_01}, 
the $\Omega_-$ mode smoothly connects to the hydrodynamic in-phase mode
[Figs.~\ref{fig:000010_di_eig_val_vec_01}(a) and~\ref{fig:000010_di_eig_val_vec_01}(c)],
and the $\Omega_+$ mode shows a bifurcation of the two purely damped modes 
[Figs.~\ref{fig:000010_di_eig_val_vec_01}(b) and~\ref{fig:000010_di_eig_val_vec_01}(e)].
Thus, changing parameters, such as the number ratio, mass ratio, and trap frequencies, induces a change of the connection between the collisionless mode and the hydrodynamic mode.
From the numerical results of the moment
equation~\eqref{eq:dipole_moment_eq_in_matrix_form},
we observe that the ratio of $M_\B\zomega_\B$ and $M_\F\zomega_\F$ is an important factor that determines the dominant species (bosons or fermions) in the collisionless mode that connects to the in-phase mode, though a rigorous mathematical proof is yet to be provided.

\begin{figure}
\includegraphics[scale=.62]{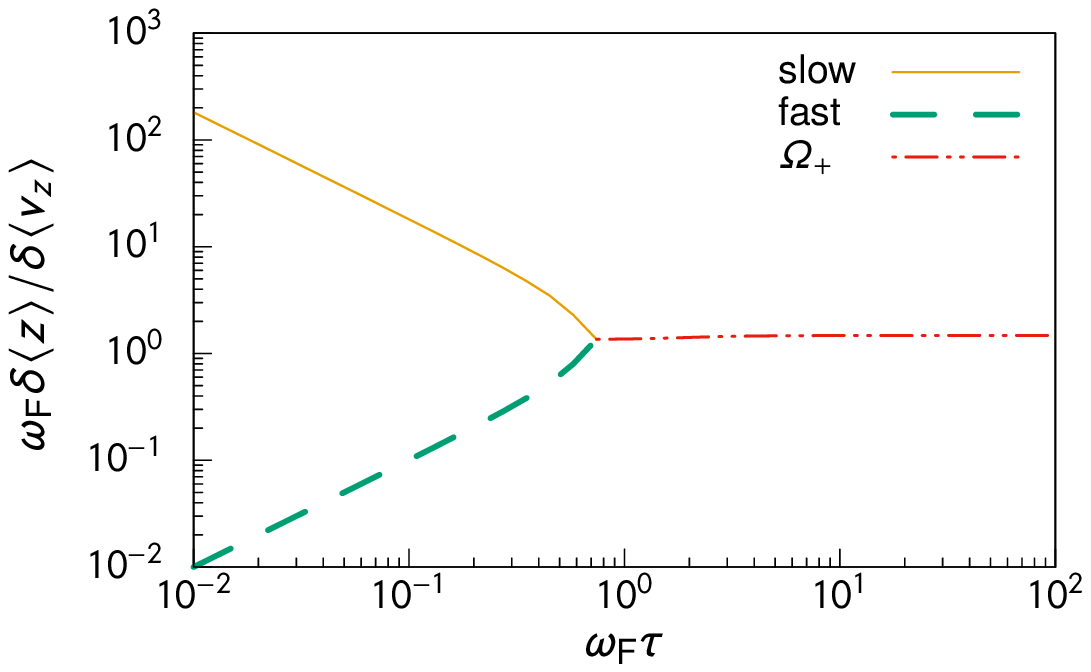}
\caption{
Ratio of the displacement of the center of mass $\delta\ev{z}$ to the velocity field $\delta\ev{v_z}$ as a function of the relaxation time for $\Delta=0$.
The transition of the $\Omega_+$ mode to both the fast and slow relaxing modes is clearly demonstrated.
The parameters are the same as those
in Fig.~\ref{fig:000010_di_eig_val_vec_01},
and we used the notations $\delta\ev{z}=\sqrt{\abs*{\delta\ev{z}_\B}^2+\abs*{\delta\ev{z}_\F}^2}$ and $\delta\ev{v_z}=\sqrt{\abs*{\delta\ev{v_z}_\B}^2+\abs*{\delta\ev{v_z}_\F}^2}$.
}
\label{fig:000010_vec_z_par_v_01}
\end{figure}

We can see
from Eqs.~\eqref{eq:eigenvector_in_hydrodynamic_fast_relaxing_mode}
and~\eqref{eq:eigenvector_in_hydrodynamic_slow_decay_mode}
that two purely damped modes have distinct features in the motion of the center-of-mass positions and the velocity fields of bosons and fermions.
In the hydrodynamic limit, the slow relaxation mode is dominated by relative displacement of the centers of mass, while the fast relaxation mode is dominated by the relative velocity of the two components.
This can be clearly seen
in Fig.~\ref{fig:000010_vec_z_par_v_01},
which shows the ratio of the displacement of the center-of-mass positions to the velocity field as a function of the relaxation time.
The ratio shows a bifurcation when the $\Omega_+$ mode turns into the two relaxation modes.
In the fast relaxing mode, the amplitude of the velocity field is much larger than that of the displacement of the center of mass, and vice versa in the slow relaxing mode.

\begin{figure}
\includegraphics[scale=.62]{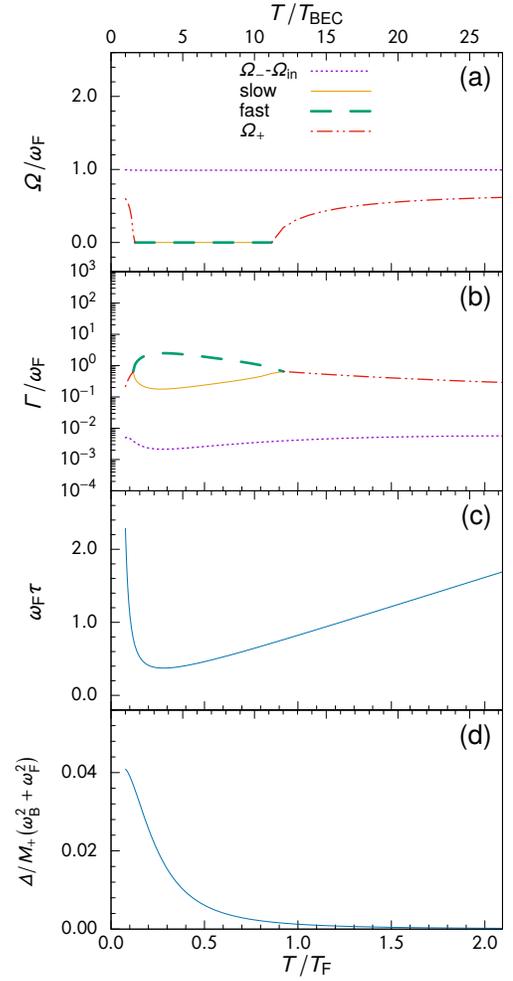}
\caption{
Temperature dependence of the dipole modes including the mean-field contribution $\Delta$ and the relaxation time $\tau$.
(a)~Frequency $\Omega$ and (b)~damping rate $\Gamma$.
(c)~Relaxation time $\tau$ and (d)~mean-field contribution $\Delta$.
We assumed a Bose--Fermi mixture of $^{87}\mathrm{Rb}$ and $^{40}\mathrm{K}$ with the same isotropic harmonic traps, giving the relations $m_\B\zomega_\B^2=m_\F\zomega_\F^2$ and $\lambda=1$.
The ratio of the total number of particles is $N_\B/N_\F=0.02$ with $N_\F\approx1.33\times10^6$ and $N_\B\approx2.67\times10^4$, and the coupling constants are $g_\BB/(l_\F^3\hbar\zomega_\F)=g_\BF/(l_\F^3\hbar\zomega_\F)\approx1.48\times10^{-1}$, which provides an $s$-wave scattering length of a Bose--Fermi collision of $a_\BF/l_\F\approx1.61\times10^{-2}$.
Here, $l_\F=\sqrt{\hbar/(m_\F\zomega_\F)}$ is the harmonic oscillator length for fermions.
To scale the temperature $T$, we use the Fermi temperature $\kB\TF=(6N_\F)^{1/3}\hbar\zomega_\F$ for a trapped ideal gas (lower horizontal axis) as well as the critical temperature of Bose--Einstein condensation $T_\BEC$ in the trapped Bose--Fermi gas (upper horizontal axis), which is determined by the condition $\mu_\B=U^0_\B(\br=\bm0)$ including the mean-field terms.
}
\label{fig:000020_val_params}
\end{figure}

\subsection{Temperature dependence}
\label{subsec:temperature_dependence}
Figures~\ref{fig:000020_val_params}(a) and~\ref{fig:000020_val_params}(b) 
show the temperature dependence of the collective modes, fully including the mean-field contribution $\Delta$ and the relaxation time $\tau$.
In calculations of 
$\Delta$
in Eq.~\eqref{eq:mean_field_contribution_in_dipole_mode}
and $\tau$ 
in Eq.~\eqref{eq:collision_rate_in_dipole_mode}, 
we use the equilibrium distribution functions including the self-consistent mean-field terms in a single-particle excitation energy.
The actual BEC critical temperature of the interacting Bose--Fermi mixture, $T_\BEC$, is determined by the condition $\mu_\B=U^0_\B(\br=\bm0)$, where $U_\B^0(\br)$ is the effective potential given by
Eq.~\eqref{eq:potential_energy_for_a_boson}
in the equilibrium state.

It is notable that the collisionless mode exists in the high-temperature region as well as the low-temperature region
[Figs.~\ref{fig:000020_val_params}(a) and~\ref{fig:000020_val_params}(b)].
The system shows a transition from the collisionless regime to the hydrodynamic regime and subsequently from the hydrodynamic regime to the collisionless regime with decreasing temperature.
Only in the intermediate temperature region $0.12\lesssim T/\TF\lesssim0.92$, which provides $\zomega_\F\tau\lesssim0.74$
[Fig.~\ref{fig:000020_val_params}(c)], 
does the collisionless mode disappear and the hydrodynamic modes emerge; the frequency of the $\Omega_+$ mode drops to zero 
[Fig.~\ref{fig:000020_val_params}(a)],
and the damping rate exhibits bifurcation to the two purely damped modes
[Fig.~\ref{fig:000020_val_params}(b)].

These behaviors are caused by the nonmonotonic temperature dependence of the relaxation time $\tau$
[Fig.~\ref{fig:000020_val_params}(c)].
In the high-temperature region, the system is in the collisionless regime with a long relaxation time, because the gas trapped in the harmonic potential becomes more dilute with increasing temperature.
The long relaxation time in the high-temperature region is in stark contrast to the uniform gas.
In the low-temperature region, the collisionless regime with a long relaxation time also emerges because of suppression of collisions due to Pauli blocking.
The effect of the mean-field contribution $\Delta$ is much smaller than that of the relaxation time $\tau$
[Fig.~\ref{fig:000020_val_params}(d)].
Ignoring the mean-field contribution $\Delta$ is a good approximation for dipole modes in the present case.
Indeed, the two purely damped modes also emerge at $\zomega_\F\tau\lesssim0.74$ in the absence of $\Delta$.

Finally, we discuss the effect of an anisotropy in the harmonic potential.
In general, this effect can be included in the self-consistent equilibrium
distributions
in Eq.~\eqref{eq:distribution_function_in_equilibrium},
the mean-field contribution $\Delta$, and the relaxation time $\tau$ through numerical calculations.
However, we can easily see the effect of the anisotropy in the special case where $\xl_\B=\xl_\F=\xl$
in Eq.~\eqref{eq:trapping_potential}.
By introducing the coordinate transformation,
\begin{align}&
\pqty\big{x',\;y',\;z'}=\pqty\big{x\xl^{1/3},\;y\xl^{1/3},\;z\xl^{-2/3}}
\;,
\label{eq:spherical_coordinate}
\end{align}
we can reduce the harmonic potential to the spherically symmetric form,
\begin{align}&
U^\trap_\xa(\br')=\frac{m_\xa\bar\zomega_\xa^2}{2}\pqty\big{x'^2+y'^2+z'^2}
\;,
\label{eq:trapping_potential_in_isotropy_form}
\end{align}
where $\bar\zomega_\xa=\xl^{2/3}\zomega_\xa$.
The calculations
of Eqs.~\eqref{eq:mean_field_contribution_in_dipole_mode}
and~\eqref{eq:collision_rate_in_dipole_mode}
can be performed after transforming to these spherically symmetric coordinates with the trap frequency $\bar\zomega_\xa$.
As a result, the mean-field contribution $\Delta$ and the relaxation time $\tau$ in the anisotropic case with $\xl\neq1$ are respectively given by $\Delta=\xl^{-4/3}\bar\Delta$ and $\tau=\bar\tau$, where $\bar\Delta$ and $\bar\tau$ are those in the isotropic case for the trap frequency $\bar\zomega_\xa$.
Although the mean-field contribution $\Delta$ is deformed from this isotropic case, the ratio $\Delta/\zomega_\xa^2=\bar\Delta/\bar\zomega_\xa^2$ is independent of the anisotropy.
In this anisotropic case,
Eq.~\eqref{eq:secular_equation_in_dipole_mode_without_anisotropy}
becomes
\begin{align}&
\pqty{\bar\omega^2-\bar\zomega_\B^2+\frac{\bar\Delta}{M_\B}}
\pqty{\bar\omega^2-\bar\zomega_\F^2+\frac{\bar\Delta}{M_\F}}
-\frac{\bar\Delta^2}{M_\B M_\F}
\notag\\&
+\xl^{2/3}\frac{\ui\bar\omega M_+}{\bar \tau}
\pqty{
\frac{\bar\omega^2-\bar\zomega_\F^2}{M_\B}
+\frac{\bar\omega^2-\bar\zomega_\B^2}{M_\F}
}
=0
\;,
\label{eq:secular_equation_in_dipole_mode_with_anisotropy}
\end{align}
where the renormalized eigenfrequency is defined as $\bar\omega=\xl^{2/3}\omega$.
The anisotropic effect is thus effectively introduced by replacing  $\tau$ with $\xl^{-2/3}\bar\tau$, with the mean-field contribution unchanged.
In the cigar, $\xl>1$, and pancake, $\xl<1$, trap cases, we obtain shorter and longer relaxation times, respectively, than in the isotropic case with $\xl=1$, and hence the hydrodynamic and collisionless center-of-mass motions are effectively enhanced by setting large and small anisotropy coefficients $\xl$, respectively.
Although this argument is based on the special case $\xl_\B=\xl_\F$, the qualitative tendency would not change in the general case $\xl_\B\neq\xl_\F$; in the cigar trap case, we obtain an effectively shorter relaxation and
thus the collisional effect is enhanced.

\subsection{Connection to experiments}
\label{discussions_with_experiments}
The temperature dependence of the dipole modes studied
in Sec.~\ref{subsec:temperature_dependence}
are specific to our trapped normal Bose--Fermi mixture.
However, the insights into dipole modes gained
from Eq.~\eqref{eq:dipole_moment_eq}
may be useful 
for other normal two-component mixtures, i.e., Bose--Bose or Fermi--Fermi mixtures with $s$-wave scattering interactions, as discussed
in Sec.~\ref{sec:method}.
An experiment on the dipole oscillation in a strongly interacting Fermi gas has been reported
in Ref.~\cite{sommer_2011_N_472_universal}.
In this experiment, two spin-polarized fermionic clouds bounce off each other and subsequently show a slow relaxation of the displacements of the two centers of mass after $160\,\mathrm{ms}$.
The bounce of the two clouds may be far-from-equilibrium dynamics and it is beyond the scope of our formalism, where we assumed a small deviation from equilibrium.
Indeed, in the hydrodynamic regime in our formalism, the bounce mode does not emerge.
On the other hand, the slow relaxation after the bounce may correspond to our slow relaxing mode in the hydrodynamic regime, the picture of which is consistent with our theoretical results, where the two separated clouds are mixed gradually.
Indeed, as discussed by using
Eq.~\eqref{eq:initial_state_to_tominate_slow_decay_mode}, 
the slow relaxing mode alone emerges in the balanced gas ($M_\B=M_\F$) with equal trap frequencies ($\zomega_\B=\zomega_\F$), where the initial two clouds are spatially separated without initial velocities.

The center-of-mass motions have also been reported in a Bose--Fermi mixture of $^7\mathrm{Li}$ and $^6\mathrm{Li}$ above and below the superfluid transition temperature~\cite{delehaye_2015_PRL_115_critical}.
In particular, phase locking is observed when both the $^7\mathrm{Li}$ and $^6\mathrm{Li}$
gases are in the normal phase;
the center-of-mass motions of the two clouds with different trap frequencies show dipole oscillations at the same frequency.
Interpreting the phase locking includes a quantum Zeno
effect~\cite{itano_1990_PRA_41_quantum}
and nonlinearities emergent from 
interspecies interactions studied in the context of
thermalization~\cite{jauffred_2017_JPBAMOP_50_universal}
based on the Caldeira--Leggett
model~\cite{ullersma_1966_P_32_exactly}.
The study in this paper has found that the phase locking of the dipole mode in a normal gas is a consequence of the hydrodynamic in-phase mode, where two gases oscillate at the same frequency given
by Eq.~\eqref{eq:frequency_of_in-phase_mode},
even if the trap frequencies of the two components are different.
In the
experiment~\cite{delehaye_2015_PRL_115_critical},
the trap frequencies of fermions and bosons are respectively given by $\omega_{z,\F}=2\uppi\times17\,\mathrm{Hz}$ and $\omega_{z,\B}\simeq2\uppi\times15.7\,\mathrm{Hz}$, and the frequency of the phase-locked dipole mode was reported to be $2\uppi\times17.9(3)\,\mathrm{Hz}$.
Since the fermionic gas is the majority ($N_\F=2.5\times10^5$ and $N_\B\sim2.5\times10^4$),  
it is natural that the frequency of the in-phase mode is likely to be relatively close to the trap frequency of the fermions, although the frequency of the hydrodynamic in-phase mode $\Omega_\mix\simeq2\uppi\times16.9\,\mathrm{Hz}$ estimated
from Eq.~\eqref{eq:frequency_of_in-phase_mode}
is, to an extent, smaller than the frequency of the phase locking reported in the
experiment~\cite{delehaye_2015_PRL_115_critical}.
The phase-locking phenomenon has been also observed in a Fermi--Fermi
mixture~\cite{gensemer_2001_PRL_87_transition}
and a Bose--Fermi
mixture~\cite{ferlaino_2003_JOBQSO_5_dipolar}
in the context of the hydrodynamic mode.

Future work will include investigating other multipole oscillations in mixtures, such as monopole and quadrupole modes accessible in experiments.
The damping process of collective modes in a Bose--Fermi mixture is also an interesting issue, including Landau
damping~\cite{shen_2015_PRA_92_landau}
and hydrodynamic damping.

\section{Conclusions}
\label{sec:conclusions}
We investigated dipole modes in a trapped normal Bose--Fermi mixture composed of single-species bosons and single-species fermions.
We theoretically obtained both the frequency and the damping rate of these modes using the moment method for the linearized Boltzmann equation in the limits of the collisionless and hydrodynamic regimes as well as in the intermediate regime between the two limits.

In the collisionless regime, there are two types of oscillating modes.
In the absence of a mean-field contribution, the frequencies of these modes correspond to the harmonic trap frequencies of the two components.
These two oscillating modes in the collisionless regime have two distinct fates in the hydrodynamic regime.
One oscillating mode turns into an in-phase oscillating mode in the hydrodynamic regime, where the frequency and the damping rate show a smooth crossover.
The other oscillation mode disappears and turns into two purely damped modes associated with the out-of-phase motion in the hydrodynamic regime, which can be regarded as a transition.
One of the purely damped modes is the fast relaxing mode, where the relative velocity of the two component is large and the gas quickly relaxes to the static equilibrium.
The other purely damped mode is the slow relaxing mode, where the two separated clouds are gradually mixed, which is unique to trapped gases.

We also studied the temperature dependence of the frequencies and damping rates of the dipole modes.
In contrast to a uniform system, trapped gases at high temperature are in the collisionless regime, because a harmonically trapped gas may spread and its density decreases with increasing temperature.
In the low-temperature region, the system may enter the collisionless regime again because of Pauli blocking in the Bose--Fermi-mixture gas.
In the intermediate-temperature region, we observe the hydrodynamic modes.
This nonmonotonic feature is highlighted when the number of bosons is smaller than the number of fermions.

Using the moment method presented in this paper, we found that dipole modes are characterized by the mean-field contribution $\Delta$ and the relaxation time $\tau$, both of which are associated with the interspecies scattering.
Even if we consider other kinds of normal two-component mixtures (Bose--Bose or Fermi--Fermi mixtures with $s$-wave scattering interactions), we can obtain the same closed set of coupled moment equations as obtained in this paper, with the relevant mean-field contribution and the relaxation time appropriate to the system of interest.
Our understanding of dipole modes obtained in this paper is thus essentially applicable to Bose--Bose or Fermi--Fermi mixtures.

\section*{Acknowledgements}
S.~W.~was supported by JSPS KAKENHI Grant No.~JP18K03499, and T.~N.~was supported by JSPS KAKENHI Grant No.~JP16K05504.
The authors wish to thank B.~Huang and R.~Grimm for their useful discussions in the earlier stage of this study.

\bibliographystyle{apsrev4-1}
\bibliography{in_harmonic_trap_pra}
\end{document}